\documentclass[twocolumn]{revtex4}
\usepackage{amssymb}
\usepackage{latexsym}
\usepackage{epsfig}
\usepackage{graphicx}
\usepackage{amsmath}
\usepackage{lipsum}
\usepackage{xcolor}
\begin{document}

\title{Some applications of the Shapiro time delay}

\author{A. Ghasemi Azar$^1$, H. Rezaei$^2$, H. Moradpour$^1$\footnote{h.moradpour@riaam.ac.ir}}
\address{$^1$ Research Institute for Astronomy and Astrophysics of Maragha (RIAAM), University of Maragheh, P.O. Box 55136-553, Maragheh, Iran\\
$^2$ Department of Mathematics, College of Sciences, Yasouj
University, Yasouj-75914-74831, Iran}

\begin{abstract}
Listening to echoes has long been a way to estimate distances, a
technique whose backbone is the time delay. The gravitational field
also creates a time delay, called Shapiro time delay, that helps
us extract some information from the field and is indeed due
to the photon journey through the field. Here, the ability of the
Shapiro effect to distinguish naked singularities from non-naked
ones (black holes) is discussed. It is also inferred that this
time delay may be hired to compare the various types of
singularities with different dimensions. Besides them, the
possibility of detecting the rotation of the assumed objects
through surveying the gravitational time delay is also addressed.
\end{abstract}

\maketitle

\section{Introduction}

Singularity is one of the attractive predictions of general
relativity (GR) \cite{Poisson, HPbook}. Although the Cosmic
Censorship Hypothesis (CCH) generally rejects the existence of
naked singularity (NS), there is not a common agreement on CCH,
and actually, NS physics has a lot to say \cite{HPbook}.
Consequently, NS formation has attracted plenty of attempts to
itself \cite{JNW, NS-2, NS-1, NS0, NS1, NS2, NS3, NS4, NS5, PRL,
NS6}. Subsequently, distinguishing NS from black holes (non-naked
singularities, i.e. singularities quarantined from the
surroundings by the event horizon) appears as a serious task for
physicists, a task that is accomplished by studying various
properties of NS \cite{d1, d2, d3, d4, d5, d6, o1, o2, o3, o4}.

Indeed, recent advances in the field of photographing black hole
candidates further encourage us to study NS and the differences
with black holes. In this regard, observational properties of
naked singularities have also been studied \cite{o1, o2, o3, o4}.
The Sagnac time delay also seems to be able to distinguish black
holes from NS \cite{d6}. In this setup, a satellite (as the
sender/receiver) orbiting the object (NS or black hole) is
crucial, a serious difficulty for the idea applicability, as such
objects, are very far from us. Therefore, another type of time
delay may be more useful for such studies, i.e. an experiment that
can be done remotely (without any need to send a satellite to long
distances).

The fourth test of general relativity, established by I. I.
Shapiro \cite{shap1}, is based on the time delay of light rays
passing through the gravitational field. In addition to being used
to check GR in the Solar system \cite{shap2, shap3, shap4}, this
effect is also helpful in verifying $i$) modified gravity theories
\cite{shapmod0, shapmod, ghosh, Dyadina:2021paa}, generalized
uncertainty principle \cite{Okcu:2021oke}, $ii$) equivalence
principle \cite{Desai:2016nqu, Boran:2018ypz, Minazzoli:2019ugi,
Kahya:2016prx}, $iii$) the number of spacetime dimensions
\cite{shapdim}, and studying the Pulsars \cite{Desai:2016nqu,
shappul0, Desai:2015hua, shappul, shappul1, shappul3, shappul2}.
Fortunately, unlike the Sagnac effect, this experiment does not
need to send a satellite and hence is possible remotely. Moreover,
although picosecond has been reported as the accuracy order of
Shapiro measurement \cite{Will:2014kxa}, it seems that the next
generation of gravitational detectors shall significantly increase
the accuracy \cite{Ballmer:2009pq, Sullivan:2020sop}. The negative
Shapiro time delay called the gravitational time advancement also
happens when the light rays passing through a weaker gravitational
field compared to the field at the observer's place, and it is
predicted that the modern versions of the Michelson-Morley
experiment shall measure this time advancement
\cite{Bhadra:2023ptq}.

Indeed, although the Schwarzschild spacetime is the most general
spherically symmetric vacuum solution and a simple solution, it is
the backbone of our understanding of many phenomena like orbits
and thermodynamics of black holes motivating physicists to study
this metric and its various generalizations
\cite{Wiltshire:2009zza,inverno,shapgamma}. Employing the
Schwarzschild metric and some generalized forms of this spacetime
including the Bardeen, Reissner-Nordstr\"{o}m, and
Ay\'{o}n-Beato-Garc\'{i}a metric, it seems that the Shapiro time
delay can distinguish these black holes from each other
\cite{Junior:2023nku}. On the other hand, to find constraints on
the deviations from spherical symmetry (SS), focusing on the
$\gamma$-metric (a non-spherically generalization of the
Schwarzschild metric that, depending on the value of $\gamma$, can
also present NS), the Shapiro time delay has been investigated
\cite{shapgamma}. The obtained time delay is equal to that of
Schwarzschild meaning that the Shapiro time delay cannot be used
to constrain $\gamma$, and hence, distinguish solutions with
different $\gamma$ (the criterion of deviation from SS).
Therefore, it seems that it is not possible to distinguish NS from
the black hole (even the simplest black hole solution i.e. the
Schwarzschild black hole) through comparing the corresponding
Shapiro time delays.

A long way has been traced to find the rotating version of the
Schwarzschild geometry called Kerr metric which also plays a vital
role in discovering the secrets of various phenomena
\cite{Wiltshire:2009zza}. Since the Schwarzschild metric is a
vacuum solution, it is then obvious to look at all other black
holes as its extensions. The Janis-Newman-Winicour (JNW) metric
and its rotational version are two well-studied generalizations of
the Schwarzschild and Kerr spacetimes, respectively \cite{JNW,d2}.
They can also include NS \cite{JNW,d2}, and attract a lot of
attention (see Refs.~\cite{JNW,d1,d2} and their citations). In
five dimensions, the Myers-Perry (MP) metric is an extension of
the Kerr black hole that reduces to the Kerr geometry whenever a
$4$-dimensional spacetime is taken into account. Correspondingly,
the $4$ and $5$-dimensional Schwarzschild metrics are also
recoverable if the zero limit of angular momentums is applied
\cite{mpmetric}. In summary, the background and importance of
these geometries requires that they be studied.

Here, our first aim is to show the power of the Shapiro test in
distinguishing black holes from NSs. To this end, the JNW
metric~\cite{JNW} and its rotating version \cite{d2} are studied
in the subsequent sections, respectively. The possibility of
verifying rotation in the fifth dimension shall be studied in the
fourth section by employing the MP metric \cite{mpmetric}. A
summary is also provided at the end.

\section{Shapiro time delay in JNW spacetime}

In the presence of the scalar field
$\Phi(=\frac{q}{2b\sqrt{\pi}}\ln(1-\frac{b}{r}))$ with charge $q$
and mass $M$, the JNW metric is obtained as

\begin{eqnarray}\label{1}
ds^2=-H^\nu
dt^2+\frac{dr^2}{H^\nu}+H^{1-\nu}r^2(d\theta^2+\sin^2(\theta)d\phi^2),
\end{eqnarray}

\noindent in which $H=1-\frac{b}{r}$, $\nu=\frac{2M}{b}$, and
$b=2\sqrt{M^2+q^2}$~\cite{JNW,d1}. The Schwarzschild spacetime is
covered at the appropriate limit of $q=0$ (or, $\nu=1$). The
spacetime represents a singularity at $r=b$ for $q\neq0$ and has
the photon sphere if $1/2<\nu\leq1$ \cite{JNW,d1}. Following the
existence and absence of the photon sphere, singularity is called
strong NS if $0\leq\nu\leq1/2$, and it is called weak NS on the
condition that $1/2<\nu<1$, respectively \cite{d1}. Indeed, it is
easy to calculate the surface area at $r=b$ through

\begin{eqnarray}\label{2}
A=\int_{\theta_0=0}^{\theta=\pi}\int_{\phi_0=0}^{\phi=2\pi}
\big[H^{1-\nu}r^2\big]_{r=b} d\theta d\phi=0,
\end{eqnarray}

\noindent for $0\leq\nu<1$ meaning that $r=b$ is a NS.

\begin{figure}[ht]
\centering
\includegraphics[width=3in, height=1in]{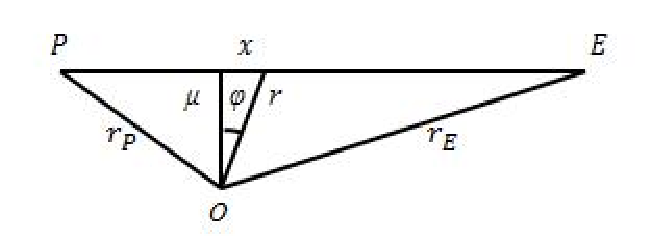}
\caption{Light ray moves towards $E$ through the path
$PE$.}
\label{fig1}
\end{figure}

Figure.~(1) displays a photon sent from point $P$ towards
$E$, while the path $PE$ has the closest vertical distance $\mu$
from the object $O$~\cite{inverno} whose surrounding spacetime is
described by the JNW metric. In this manner, for a small
displacement $dx$, we have

\begin{eqnarray}\label{3}
\bigg\{^{dr=\frac{x}{\sqrt{x^2+\mu^2}}dx}_{d\phi=\frac{\mu}{x^2+\mu^2}dx}\Rightarrow
d\phi=\frac{\mu}{r\sqrt{r^2-\mu^2}}dr.
\end{eqnarray}

\noindent Thus, since for a light ray ($ds^2=0$) on the
$\theta=\frac{\pi}{2}$ plane

\begin{eqnarray}\label{4}
H^\nu dt^2=\frac{dr^2}{H^\nu}+H^{1-\nu}r^2d\phi^2,
\end{eqnarray}

\noindent one can easily use Eq.~(\ref{3}) to find
%
%
%


\begin{eqnarray}\label{5}
dt^2=H^{-2\nu}\big[1+\frac{\mu^2}{r^2-\mu^2}H\big]dr^2,
\end{eqnarray}

\noindent as the time delay of a photon passing the path of $PE$ in
the presence of an object whose spacetime is described by the JNW
metric. At long distances from $O$, one can approximate

\begin{eqnarray}\label{6}
dt^2\simeq\frac{r^2}{r^2-\mu^2}\big[1+\frac{2b}{r}\nu-\frac{b\mu^2}{r^3}-\frac{2b^2\mu^2}{r^4}\nu\big]dr^2,
\end{eqnarray}

\noindent leading to


\begin{eqnarray}\label{7}
dt\simeq\frac{rdr}{\sqrt{r^2-\mu^2}}\bigg[1+\frac{2M}{r}-\frac{\mu^2}{r^2}(\frac{1}{r}+\frac{4M}{r^2})\sqrt{M^2+q^2}\bigg],
\end{eqnarray}

\noindent where assuming $\frac{b}{r}\ll1$, the Taylor expansion
has been used (for example, we have $H^{-2\nu}\simeq1+\frac{2\nu
b}{r}$) and terms including powers upper than $(\frac{1}{r})^4$
are ignored. In fact, unlike Ref.~\cite{inverno}, here, the
$(\frac{1}{r})^4$ term is calculated and the Schwarzschild result
is also obtainable by adopting $q=0$ (or equally, $\nu=1$)
\cite{inverno}. Accordingly, the discrepancy with the
Schwarzschild case emerges in terms including $q$. It is apparent
that $dt$ decreases as $q$ increases meaning that, for the same
$dr$, NS produces less $dt$ compared to the Schwarzschild black
hole, and accordingly, those NSs that have the photon sphere,
generate more time delay rate compared to those without the photon
sphere.

A Shapiro time delay measurement helps us confine and estimate the
value of $q$. To show it, up to the first order of expansion, one
can write Eq.~(\ref{7}) as $dt\sim dt_{S}-q^2\delta t$, where
$dt_S\equiv dt\big|_{q=0}$ (the Schwarzschild case) and $\delta
t\equiv\frac{\mu^2dr}{2Mr^2\sqrt{r^2-\mu^2}}(1+\frac{4M}{r})$.
Now, consider an object with mass $M$ and a measurement with
uncertainty $A$ reporting $t=\int dt$ for the time delay. If the
object is supposed to be a Schwarzschild black hole, then
mathematical calculations give us $t_S=\int dt_S$. In this manner,
if $|t-t_S|<A$, then it can be said that with the accuracy $1-A$,
the object is a Schwarzschild black hole.

On the other hand, $|t-t_S|<A$ can also be used to find an upper
bound on the value of $q$ as $q^2<\frac{A}{\int \delta t}$.
Therefore, by increasing the precision of the setup, one can find
more accurate upper bound on $q$ (or equally, one can find the
value of $q$ with more certainty). In this line, it is worthy to
mention that although the current detectors have also significant
precision i.e. $A\sim O\big(10^{-12}\big)$s \cite{Will:2014kxa},
the next generation of gravitational wave detectors equips us with
more accurate measurements \cite{Ballmer:2009pq,
Sullivan:2020sop}.

\section{Rotating $4$-dimensional metric}

The geometry of spacetime including a rotating object (the
rotational version of the JNW metric) is described as \cite{d2}

\begin{eqnarray}\label{8}
&&ds^2=-h^{1-\nu}\rho(\frac{dr^2}{\Delta}+d\theta^2+\sin^2\theta
d\phi^2)+\\&&h^\nu(dt-a\sin^2\theta
d\phi)^2+2a\sin^2\theta(dt-a\sin^2\theta d\phi)d\phi,\nonumber
\end{eqnarray}

\noindent where

\begin{eqnarray}\label{9}
&&h=1-\frac{br}{\rho},\nonumber\\
&&\nu=\frac{M}{\sqrt{M^2+q^2}}=\frac{2M}{b},\nonumber\\
&&\Delta=r^2+a^2-br,\nonumber\\
&&\rho=r^2+a^2\cos^2\theta,
\end{eqnarray}

\noindent and the scalar field $\Phi$ takes the form

\begin{eqnarray}\label{10}
\frac{q}{2b}\ln(1-\frac{br}{\rho}).
\end{eqnarray}

\noindent In the above expressions, $q$ and $a=J/M$ are the scalar
charge and angular momentum per mass, respectively, while $J$
denotes the angular momentum and the JNW case is easily recovered
at the $J=0$ limit. It is reduced to the Kerr and Schwarzschild
solutions for $q=0(\equiv \nu=1)$ and $a=q=0$, respectively. For
other values of $\nu$ ($0<\nu<1$), the line element~(\ref{8})
includes a naked singularity \cite{d2}.

Now, considering the approach of the previous section, one reaches

\begin{eqnarray}\label{11}
&&h^\nu dt^2+\frac{2\mu a}{r\sqrt{r^2-\mu^2}}(1-h^\nu)drdt+\\
&&[\frac{\mu^2a^2(h^\nu-2)}{r^2(r^2-\mu^2)}-(\frac{\mu^2}{r^2(r^2-\mu^2)}+\frac{1}{\Delta})h^{1-\nu}\rho]dr^2=0.\nonumber
\end{eqnarray}

\noindent for the photon moving on the plane $\theta=\pi/2$, and
easily, one can recover Eq.~(\ref{5}) at the appropriate limit
$a=0$ leading to $h=H$. Of course, whenever $\theta=\pi/2$, we
always have $h=H$ ($\rho=r^2$) and thus

\begin{eqnarray}\label{12}
&&\frac{2\mu a}{r\sqrt{r^2-\mu^2}}(h^{-\nu}-1)\simeq\frac{4M\mu
a}{r^2\sqrt{r^2-\mu^2}}\equiv-\Theta,\nonumber\\&&\Delta\simeq
r^2+a^2,\\
&&(\frac{\mu^2h^{1-2\nu}\rho}{r^2(r^2-\mu^2)}+\frac{h^{1-2\nu}\rho}{\Delta})\simeq\nonumber\\
&&(1-\frac{b\mu^2}{r^3}-\frac{(r^2-\mu^2)a^2}{r^2(r^2+a^2)})\frac{r^2h^{-2\nu}}{r^2-\mu^2},\nonumber\\
&&\frac{\mu^2a^2(h^\nu-2)}{h^\nu
r^2(r^2-\mu^2)}\simeq\nonumber\\
&&-\frac{\mu^2a^2}{r^2(r^2-\mu^2)}(1+\frac{2\nu
b}{r})\simeq-\frac{\mu^2a^2}{r^2(r^2-\mu^2)}h^{-2\nu},\nonumber
\end{eqnarray}


\noindent whenever $\frac{b}{r}\ll1$. It finally gives the
coefficient of $dr^2$ as

\begin{eqnarray}\label{13}
&&-\frac{h^{-2\nu}r^2}{r^2-\mu^2}\big(1-\frac{b\mu^2}{r^3}-f(a)\big)\simeq-\frac{r^2}{r^2-\mu^2}
\bigg[-\frac{b\mu^2}{r^3}\nonumber\\&&(1-f(a))(1+\frac{2b}{r}\nu)-\frac{2b^2\mu^2}{r^4}\nu\bigg]\equiv-\alpha,\nonumber\\
&&f(a)=\frac{\frac{a^2}{r^2}[1-\frac{\mu^2}{r^2}(2+\frac{a^2}{r^2})]}{1+\frac{a^2}{r^2}},
\end{eqnarray}

\noindent compared to Eq.~(\ref{6}) to see that, at this level of
approximation, the effects of $a$ are stored into $f(a)$.
Therefore, at this limit, Eq.~(\ref{11}) takes the form
$dt^2-\Theta drdt-\alpha dr^2\simeq0$ (the alternative of
Eq.~(\ref{6})) that eventually renders

\begin{eqnarray}\label{14}
&&dt\simeq\bigg[\sqrt{\alpha}+\frac{\Theta(\frac{\Theta}{4\sqrt{\alpha}}+1)}{2}\bigg]dr,\nonumber\\
&&\sqrt{\alpha}\simeq\frac{r}{\sqrt{r^2-\mu^2}}\bigg[1-\frac{f(a)}{2}+(1-f(a))\frac{2M}{r}\nonumber\\&&-\frac{\mu^2}{r^2}(\frac{1}{r}+\frac{4M}{r^2})\sqrt{M^2+q^2}\bigg],
\end{eqnarray}

\noindent as the counterpart of Eq.~(\ref{7}) whenever $a\neq0$
and of course provided that $f(a)\ll1$. The plausibility of the
latter condition is a reflection of our great distance from $O$ in
agreement with $i)$ the primary assumption $b/r\ll1$ and also,
$ii)$ the fact that such objects are very far from us. Indeed, it
is the only solution of $dt^2-\Theta drdt-\alpha dr^2\simeq0$ that
produces Eq.~(\ref{7}) at the limit of $a=0$. Thus, the ability of
the Shapiro time delay in detecting rotation in $4$-dimensional
spacetime as well as distinguishing rotating NS from the rotating
black hole is deduced.

\section{The MP spacetime}

One of the Universe's mysteries is its number of dimensions which
seems searchable using the Shapiro time delay~\cite{shapdim}.
Another puzzle is the method of detecting motions in higher
dimensions when we do not have direct access to the higher
dimensions. Here, focusing on the MP metric, we are going to
provide an answer by studying the effects of such a movement on
the Shapiro time delay. The MP geometry is \cite{mpmetric}

\begin{eqnarray}\label{15}
&&ds^2=-dt^2+\frac{M(dt+a\sin^2\theta d\phi+\beta\cos^2\theta
d\psi)^2}{r^2+a^2\cos^2\theta+\beta^2\sin^2\theta}\nonumber\\&&
+\frac{r^2(r^2+a^2\cos^2\theta+\beta^2\sin^2\theta)}{(r^2+a^2)(r^2+\beta^2)-Mr^2}dr^2\\&&+(r^2+a^2)\sin^2\theta
d\phi^2+(r^2+\beta^2)\cos^2\theta
d\psi^2\nonumber\\&&+(r^2+a^2\cos^2\theta+\beta^2\sin^2\theta)d\theta^2,\nonumber
\end{eqnarray}

\noindent where $\beta$ is related to the angular momentum of
fifth dimension.

For $\beta=a=0$, the five-dimensional Schwarzschild solution, i.e.

\begin{eqnarray}\label{16}
&&ds^2=-(1-\frac{M}{r^2})dt^2+\frac{dr^2}{1-\frac{M}{r^2}}\\&&+r^2(d\theta^2+\sin^2\theta
d\phi^2+\cos^2\theta d\psi^2),\nonumber
\end{eqnarray}

\noindent is achieved. Following the approach of the previous section,
one finds

\begin{eqnarray}\label{17}
dt^2=(1-\frac{M}{r^2})^{-2}\big[1+\frac{\mu^2}{r^2-\mu^2}(1-\frac{M}{r^2})\big]dr^2,
\end{eqnarray}

\noindent as the Shapiro time delay of a $5$-dimensional
Schwarzschild black hole whenever $\theta=\pi/2$ and $d\psi=0$. At
long distances from $O$ (or equally, $M/r\ll1$), one easily reaches

\begin{eqnarray}\label{18}
&&dt^2\simeq\frac{r^2}{r^2-\mu^2}[1+\frac{2M}{r^2}-\frac{M\mu^2}{r^4}]dr^2,\\
&&\Rightarrow
dt\simeq\frac{rdr}{\sqrt{r^2-\mu^2}}[1+\frac{M}{r^2}-\frac{M\mu^2}{2r^4}],\nonumber
\end{eqnarray}

\noindent that clearly explains if the Shapiro time delay of a
Schwarzschild candidate obeys this equation instead of the $q=0$
case of Eq.~(\ref{7}), then one can claim that the Universe has
$5$ dimensions. Compared to Eq.~(\ref{7}), it is seen that $1/r^3$
does not appear here, a fact that originated from the $M/r^2$ term in
the metric~(\ref{17}) (for the Schwarzschild case, we have $M/r$).
Moreover, it should be noted that the origin of $r^{-4}$ also
differs from that of this term in Eq.~(\ref{7}) as their
coefficients differ. Finally and as a preliminary check, if we
only want to keep the first two terms $\big(1+\frac{M}{r^2}$ in
this equation, and $1+\frac{2M}{r}$ in Eq.~(\ref{7})$\big)$, then
it is enough to replace $M/r^2$ and $2M/r$ with each other to see
that the corresponding time delays are mutually recovered, a net
reflection of the relationship between the corresponding metrics
(Eqs.~(\ref{5}) and~(\ref{17})).

A three dimensional subspace of metric~(\ref{15}) with
$\theta=\pi/2$ and $\psi=0$ gives the geometry

\begin{eqnarray}\label{19}
&&ds^2=-(1-\frac{M}{r^2+\beta^2})dt^2+\frac{2Ma}{r^2+\beta^2}dtd\phi\\&&
+\frac{r^2(r^2+\beta^2)}{(r^2+a^2)(r^2+\beta^2)-Mr^2}dr^2\nonumber\\&&+(r^2+a^2+\frac{Ma^2}{r^2+\beta^2})d\phi^2,\nonumber
\end{eqnarray}

\noindent leading to

\begin{eqnarray}\label{20}
&&0=-(1-\frac{M}{r^2+\beta^2})dt^2+\frac{2Ma\mu}{r(r^2+\beta^2)\sqrt{r^2-\mu^2}}dtdr\nonumber\\&&+
\big[(r^2+a^2+\frac{Ma^2}{r^2+\beta^2})\frac{\mu^2}{r^2(r^2-\mu^2)}\nonumber\\&&+\frac{r^2}{r^2+a^2-\frac{Mr^2}{r^2+\beta^2}}\big]dr^2,
\end{eqnarray}

\noindent when one uses Eq.~(\ref{3}) and considers a photon
($ds=0$). Clearly, Eq.~(\ref{17}) is produced for $a=\beta=0$, and
additionally, even if the object $M$ does not rotate in $4$
dimension ($a=0$), then the existence of the fifth dimension
rotation still contributes to the results on the condition that
$\beta\neq0$. Hence, the detection of rotation in higher
dimensions through the Shapiro effect is possible. When $a=0$ and
$\beta\neq0$, the solution is

\begin{eqnarray}\label{21}
dt\simeq\frac{rdr}{\sqrt{r^2-\mu^2}}[1+\frac{M}{r^2+\beta^2}-\frac{M\mu^2}{2r^2(r^2+\beta^2)}],
\end{eqnarray}

\noindent where $\frac{M}{r^2+\beta^2}\ll1$ has been assumed.
Therefore, rotation in the fifth dimension is verifiable using
this effect. Clearly, it also reduces to Eq.~(\ref{18}) when
$\beta=0$. For the $\beta=0$ case (when $a\neq0$), as the
coefficient of $dtdr$ in Eq.~(\ref{20}) is not zero, following the
approach that led to Eq.~(\ref{14}), one finally obtains
$dt^2-\varphi drdt-\epsilon dr^2\simeq0$ yielding

\begin{eqnarray}\label{22}
&&dt\simeq\bigg[\sqrt{\epsilon}+\frac{\varphi(1+\frac{\varphi}{4\sqrt{\epsilon}})}{2}\bigg]dr,\\
&&\sqrt{\epsilon}\simeq\frac{r}{\sqrt{r^2-\mu^2}}\bigg[1-\frac{F(a)}{2}+(1-F(a))\frac{M}{r^2}\nonumber-\frac{M\mu^2}{2r^4}\bigg],
\end{eqnarray}

\noindent as the only solution that recovers Eq.~(\ref{18}) at the
limit $a=0$. Here, $\varphi=\frac{2Ma\mu}{r^3\sqrt{r^2-\mu^2}}$
and

\begin{eqnarray}\label{23}
F(a)=\frac{a^2}{r^2}\bigg(\frac{1-\frac{\mu^2}{r^2}\big[2+\frac{a^2}{r^2}-\frac{M}{r^2}\big]}{1+\frac{a^2}{r^2}-\frac{M}{r^2}}\bigg),
\end{eqnarray}

\noindent and thus $\varphi=F(a)=0$ for $a=0$.

Generally, when $a,\beta\neq0$, long calculations lead to
$dt^2-\psi drdt-\mathcal{N} dr^2\simeq0$ and thus

\begin{eqnarray}\label{24}
dt\simeq\bigg[\sqrt{\mathcal{N}}+\frac{\psi(1+\frac{\psi}{4\sqrt{\mathcal{N}}})}{2}\bigg]dr,
\end{eqnarray}

\noindent where
\begin{eqnarray}\label{25}
\sqrt{\mathcal{N}}&\simeq&\frac{r}{\sqrt{r^2-\mu^2}}\bigg[1-\frac{\mathcal{F}(a)}{2}+\frac{M(1-\mathcal{F}(a))}{r^2+\beta^2}\nonumber\\&-&\frac{M\mu^2}{2r^2(r^2+\beta^2)}\bigg],\nonumber\\
\psi&=&\frac{2Ma\mu}{r(r^2+\beta^2)\sqrt{r^2-\mu^2}},\\
\mathcal{F}(a)&=&\frac{a^2}{r^2}\bigg(\frac{1-\frac{\mu^2}{r^2}\big[2+\frac{a^2}{r^2}-\frac{M}{r^2+\beta^2}\big]}{1+\frac{a^2}{r^2}-\frac{M}{r^2+\beta^2}}\bigg).\nonumber
\end{eqnarray}

\noindent Clearly, Eqs.~(\ref{21}) and~(\ref{22}) are obtained at
the appropriate limits $a=0$ and $\beta=0$, respectively. Of
course, comparing Eqs.~(\ref{21}) and~(\ref{18}), it is understood
that one could have achieved this result by replacing $M/r^2$ with
$\frac{M}{r^2+\beta^2}$ in Eqs.~(\ref{22}) and~(\ref{23}).

In summary, while Eq.~(\ref{21}) implies on the implication of the
rotation in the fifth dimension of the Shapiro time delay,
difference between $F(a)$ and $f(a)$ clearly shows that even the
existence of the fifth dimension affects the time delay
($\beta=0$). Indeed, as it is emphasized by the information stored in $\mathcal{F}(a)$, these effects become more tangible when $\beta\neq0$.
This achievement is strengthened by the point mentioned after
Eq.~(\ref{18}), where the emergence of $M/r^2$ instead of $M/r$ in
the time delay of a five-dimensional Schwarzschild spacetime has
been argued, a feature again confirmed by comparing
Eqs.~(\ref{22}) and~(\ref{14}) with each other.

\section{Conclusion}

Depending on the metric, it is possible to distinguish NSs from
black holes via the gravitational time delay (the Shapiro effect).
Moreover, the possibility of verifying the existence of extra
dimensions and detecting rotation in higher dimensions
through the Shapiro time delay has also been studied. The results
imply the ability of this effect in such investigations. Indeed,
comparing Secs.~($\textmd{III}$) and~($\textmd{IV}$) shows that
this effect can even be used to compare objects of various
dimensions.

Therefore, we can hope that the use of the Shapiro time delay and
such fundamental experiments shall help us analyze the dimensions
of the universe and its contents. Finally, it is worthwhile to
mention that challenging the ideas presented here by different
data such as GW could be an interesting topic for future projects.
\section*{Declarations Conflict of interest}
The authors declare no conflict of interest.
\section*{Acknowledgments} The authors would like to appreciate
the anonymous referee for the valuable comments.

\end{document}